\documentclass[sigplan,screen]{acmart}
\usepackage{xurl}
\usepackage{array}
\usepackage{enumitem}
\usepackage{float}

\newcommand{\cmd}[1]{\texttt{\detokenize{#1}}}
\newcommand{\toolname}{RepoTrace}

\AtBeginDocument{%
}

\setcopyright{cc}
\setcctype{by-nc-nd}
\acmDOI{10.1145/3837729.3840498}
\acmYear{2026}
\copyrightyear{2026}
\acmISBN{979-8-4007-2910-2/2026/10}
\acmConference[SPLASH Companion '26]{Companion Proceedings of the 2026 ACM SIGPLAN International Conference on Systems, Programming, Languages, and Applications: Software for Humanity}{October 3--9, 2026}{Oakland, CA, USA}
\acmBooktitle{Companion Proceedings of the 2026 ACM SIGPLAN International Conference on Systems, Programming, Languages, and Applications: Software for Humanity (SPLASH Companion '26), October 3--9, 2026, Oakland, CA, USA}
\acmSubmissionID{splashcomp26demo-p39-p}
\received{2026-06-27}
\received[accepted]{2026-07-25}

\begin{document}

\title{\toolname{}: Browser-Assisted Evidence Collection for GitHub Research Datasets}

\author{Xue Yao}
\orcid{0009-0001-5818-5912}
\affiliation{%
  \institution{Monash University}
  \city{Melbourne}
  \country{Australia}}
\email{xyao0028@student.monash.edu}

\author{Zehua Zhang}
\orcid{0009-0003-2132-0361}
\affiliation{%
  \institution{Monash University}
  \city{Melbourne}
  \country{Australia}}
\email{zzha0610@student.monash.edu}

\author{Jiatong Liu}
\orcid{0009-0000-5040-8272}
\affiliation{%
  \institution{Monash University}
  \city{Melbourne}
  \country{Australia}}
\email{jliu0455@student.monash.edu}

\author{Yongqiang Tian}
\correspondingauthor
\orcid{0000-0003-1644-2965}
\affiliation{%
  \institution{Monash University}
  \city{Melbourne}
  \country{Australia}}
\email{yongqiang.tian@monash.edu}

\begin{abstract}
Empirical software engineering studies frequently build datasets from GitHub
issues and pull requests. In many projects, researchers inspect pages in a
browser, copy selected fields into spreadsheets, keep side notes in separate
documents, and later run scripts to normalize or export the data. This workflow
is flexible, but the page evidence, the research codes, and the rationale behind
each decision end up spread across tabs and files, making provenance, update
tracking, and multi-reviewer labeling difficult to audit.

\toolname{} is a browser-assisted research tool that collects GitHub issue and
pull-request evidence into a local SQLite-backed workspace. It combines a Chrome
side-panel extension, an Express backend, and a React dashboard to capture page
snapshots, comments, labels, notes, screening and labeling decisions, refresh
history, and scoped exports, keeping the source evidence and the research
interpretation linked together.

A validation pass covered 20 Matplotlib issues across two study projects. The
resulting dataset preserves 22 snapshots, 38 comments, 20
research notes, 98 annotations, 20 screening reviews, 20 fix-evidence entries,
and 4 simulated unresolved consensus conflicts. The results show that
\toolname{} can support a complete local evidence-collection workflow for
manually constructed GitHub issue and pull-request datasets.
\end{abstract}



\maketitle

\section{Introduction}

Many empirical software engineering studies begin by identifying relevant
GitHub issues or pull requests, inspecting their discussion, deciding whether
each belongs in the study, and assigning research labels. In practice, this
workflow is iterative and evidence-heavy. Researchers read
issue bodies, comments, linked pull requests, patches, tests, and maintainer
labels, and they must revisit records as those records evolve. They also need to
record why a record was included or excluded, which evidence supports each
label, and where reviewers disagree. These needs are common in
repository-mining research, which routinely draws on evolving socio-technical
evidence recorded on GitHub, such as pull-based development
traces and contribution-evaluation signals
\cite{kalliamvakou2014promises,gousios2014exploratory,tsay2014influence}.
Curated issue datasets also underpin bug studies across domains, including deep
learning compilers~\cite{DBLP:conf/sigsoft/ShenM0TCC21}, the Solidity
compiler~\cite{DBLP:conf/issta/MaZS00C24}, and data-visualization
libraries~\cite{DBLP:journals/pacmse/LuTZMXCS25}.

Two common approaches leave parts of this process unsupported. Spreadsheet-based
coding is lightweight and flexible: a row can hold a URL, a few labels, and a
note. With enough discipline a researcher can also paste screenshots or archive
pages by hand, so the limitation is not that any single field is impossible to
store. The difficulty is keeping the page evidence, the evolving research codes,
and the record of \emph{why} a decision was made linked together and queryable
as the study progresses, rather than scattered across tabs, files, and ad-hoc
columns. Mining scripts collect structured data at scale
but are typically disconnected from the manual reading and qualitative coding
that such datasets depend on, and they discard the rendered page context a
reviewer later wants to re-examine. This separation leaves a gap between the source
artifact, the researcher's evolving interpretation, and the exported dataset
used in a paper.

\toolname{} provides a local research workbench for browser-first
GitHub data collection, aimed at researchers and students who construct issue
and pull-request datasets manually or semi-manually. Rather than replacing
researcher judgment, the tool stores evidence and labels together so that the
judgment remains auditable. \toolname{} treats a GitHub
issue or pull request as both a source document and a changing research object,
storing the browser-collected snapshot, extracted metadata, research labels,
notes, review decisions, and refresh history in one local workspace. This design
follows prior warnings that mined repository data must be interpreted with care,
because repository traces may omit context or encode process assumptions
\cite{kalliamvakou2014promises,barros2021cookbook}.

\toolname{} combines a Chrome side panel, a SQLite-backed backend, a React
dashboard, scoped JSON exports, project backup, and refresh tracking through five
capabilities:
\begin{itemize}[topsep=0pt, leftmargin=*]
  \item \textbf{Evidence capture.} The side panel collects selected page fields,
  comments, labels, snapshots, and provenance while a researcher reads a GitHub
  page. The original evidence stays next to each coding decision, reducing
  manual copying and pasting.
  \item \textbf{Project-scoped organization.} Each project keeps source metadata
  separate from research labels, notes, screening decisions, linked artifacts,
  and fix evidence, which makes exploratory collections and final datasets easier
  to maintain.
  \item \textbf{Review and labeling workflow.} A single dashboard lets
  researchers filter records, inspect evidence, assign labels, check data
  quality, and surface disagreements without switching between browser tabs,
  spreadsheets, and notes.
  \item \textbf{Refresh and audit history.} Refresh checks reveal whether
  collected records changed after initial inspection and preserve an audit trail
  of later updates, so stale evidence is caught before export.
  \item \textbf{Export and backup.} Scoped exports produce one record, a
  filtered dataset, or a complete project backup for paper analysis, collaborator
  review, and artifact sharing.
\end{itemize}

We demonstrate \toolname{} on a self-collected dataset of 20
\texttt{matplotlib/matplotlib} issues organized into two study projects,
spanning backend/rendering defects and regression/compatibility evidence. The
walkthrough exercises the full workflow---collection, screening, labeling,
consensus review, refresh, and scoped export---and shows that the tool preserves
the page snapshots, comments, GitHub labels, research labels, notes, and
rationales on which the coding decisions depend. \toolname{} is released as a
reproducible source artifact with seeded demo data and an automated test suite.
A demonstration video accompanies the artifact.

\section{Background and Related Work}

\toolname{} relates to two classes of existing tools. \emph{Repository-mining
frameworks} collect GitHub data at scale: GHTorrent mirrors GitHub's event
stream into a queryable dataset \cite{gousios2013ghtorrent}, PyDriller offers a
Python API over commit and change history \cite{spadini2018pydriller}, and the
GitHub REST and GraphQL APIs provide similar structured access
\cite{githubrestapi}. Boa provides a language and infrastructure for
ultra-large-scale repository and source-code
mining~\cite{DBLP:journals/tosem/0001NRN15}, while
SEART Data Hub constructs and preprocesses large source-code datasets from
public GitHub repositories~\cite{DBLP:conf/icsm/DabicTB24}. These tools are
designed for broad repository mining,
but they return normalized records rather than the rendered page a reviewer
reads, and they are decoupled from the manual screening and qualitative coding
that researcher-curated datasets rely on. \emph{Qualitative data-analysis tools}
such as MAXQDA, ATLAS.ti, and NVivo support codebooks, multi-rater coding, and
inter-rater agreement, but they are document-centric and do not integrate with
GitHub: they
do not detect issue or pull-request pages, extract their source metadata, or
track when the underlying artifact changes after coding. Researchers
therefore bridge the two with spreadsheets and ad-hoc scripts, recreating the
very provenance gap that prior mining-methodology work cautions against
\cite{kalliamvakou2014promises,barros2021cookbook}.

General data-management systems address a different stage of the research
lifecycle. YARD standardizes archival curation of research outputs
\cite{DBLP:journals/datascience/PeerD20}, while DataLad supports distributed
data versioning, provenance, and
collaboration~\cite{DBLP:journals/jossw/HalchenkoMPSWGM21}. Their focus is on
dataset packaging and reuse rather than on capturing record-level GitHub
evidence and linking it to evolving qualitative decisions during dataset
construction.

\toolname{} instead focuses on manual dataset construction. Rather than mining
at scale or coding generic documents, it couples browser-based evidence capture
that preserves the rendered page with study-specific labeling, screening,
refresh tracking, and an audit trail in a single local workspace. Its scope is
manual, qualitative dataset
construction, where the live evidence, the evolving codes, and the rationale for
each decision need to stay linked together throughout the study.

\section{Tool Design and Workflow}

\begin{figure}[htbp]
  \centering
  \includegraphics[width=\columnwidth]{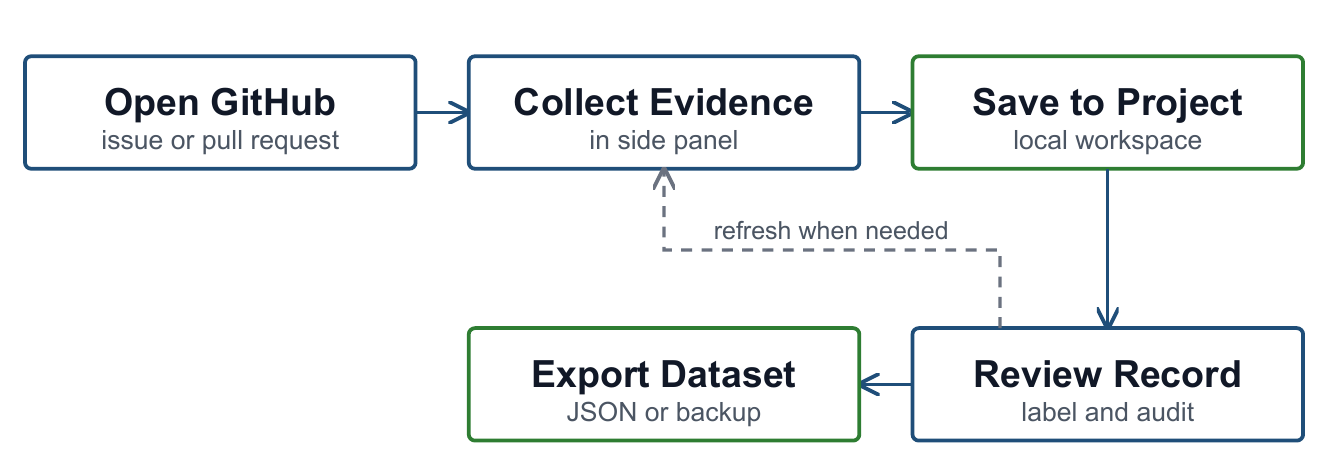}
  \caption{\toolname{} user workflow, from browser-based collection through
  dashboard review and labeling, refresh and audit, to scoped export and
  backup.}
  \label{fig:workflow}
\end{figure}

\toolname{} is organized around dataset projects. A project represents a single
study, such as a study of bugs or backend behavior in a specific repository.
Every collected record
belongs to one project, so researchers can separate exploratory collections,
final datasets, and demo data. As summarized in Figure~\ref{fig:workflow}, the
workflow follows five steps: open a GitHub
issue or pull request, collect evidence in the side panel, review and label the
record in the dashboard, refresh or audit it when needed, and export a JSON
dataset or project backup.

\subsection{Collection from GitHub Pages}

The Chrome extension detects supported GitHub issue and pull-request pages. In
the side panel, the researcher inspects the detected repository, record number,
title, URL, and duplicate status for the current project, then chooses which
visible attributes to preserve, such as title, body text, labels, state,
timestamps, comments, and the raw snapshot. Before saving the record to the
local backend, the side panel can also attach a current-page note, research
labels, a screening decision, fix evidence, and manually linked artifacts.

Browser extraction is the primary collection path. Optional GitHub API support
can enrich or refresh records, but the workflow requires no token, which keeps
\toolname{} usable in local classroom and artifact-review settings. Browser-first collection also preserves discussion
context that API responses may normalize or omit, including visible timeline
text, inline links, and page-specific evidence. Issue discussions and bug
reports often contain the reproduction steps, expected and observed behavior,
environment details, and follow-up clarification needed to judge the quality of
a record \cite{junior2021labelitbe,yang2023ossaiissues}.

\subsection{Research Dashboard}

The dashboard provides project overviews, record search, queue cards, filters,
record details, research-label management, notes, annotations, fix evidence,
linked artifacts, consensus state, quality checks, audit logs, and refresh
history. Researchers can filter by source labels, research labels, reviewers,
inclusion decisions, fix status, fix strategy, linked artifacts, and evidence
completeness. Workflow queue cards surface records that need a
second review, consensus, taxonomy backfill, update review, or refresh.

Record details are organized into three views. The evidence view shows collected
source fields, snapshots, comments, timeline text, and provenance; the research
view shows notes, labels, screening rationale, fix evidence, consensus
candidates, and backfill decisions; and the log view shows audit and change
history.

\subsection{Research Labels and Taxonomy Evolution}

\toolname{} distinguishes GitHub labels from research labels. GitHub labels are
source metadata maintained by the project, such as \texttt{bug},
\texttt{Documentation}, or \texttt{status: needs clarification}. Research labels
are study-specific codes created by the researchers, such as symptom, component,
evidence source, fix status, confidence, or fix strategy. The two play different
roles: a maintainer label is evidence about how the project classified an issue,
whereas a research label is an analytical decision made under the study protocol.

Research-label categories can evolve during analysis. When a user adds a new
category or label, \toolname{} records a taxonomy-change entry and queues
backfill review for older records that may need reconsideration, preventing
silent relabeling and keeping the historical coding process auditable. For
multi-reviewer coding, \toolname{} stores the labeler, confidence, rationale, and
status of each annotation, and surfaces records with conflicting labels under
the same category for consensus review. These disagreement records can later
support standard inter-rater reliability and agreement checks when a study
protocol requires quantitative reliability reporting
\cite{diaz2021irr,hoda2021stgt}.

\subsection{Exports for Research Use}

\toolname{} distinguishes three export scopes. A single-record export creates a
reviewable evidence package for one record. A filtered-record export captures
the current dashboard result set for analysis after applying project,
repository, label, or workflow filters. A project backup
exports a complete project for restoration or artifact sharing. Keeping these
scopes distinct avoids conflating a full database backup with a research dataset
or a single-case evidence package.

\subsection{Refresh and Change Tracking}

GitHub records can change after collection. Maintainers may close the issue,
add labels, edit the description, or post new
comments. \toolname{} therefore provides a single-record refresh and a
project-level \textit{Check Updates} action; a project-level run summarizes
changed, unchanged, and failed records. When changes are detected, \toolname{}
stores a new snapshot, records change-log entries, and marks the record as
needing update review. Because the tool is local-first, it does not continuously
synchronize with GitHub, but it gives researchers an
explicit checkpoint before exporting or submitting a dataset. The optional API
refresh path uses GitHub's documented REST API as an enrichment channel rather
than as the sole source of study evidence \cite{githubrestapi}.

\section{Implementation}

\begin{figure}[htbp]
  \centering
  \includegraphics[width=\columnwidth]{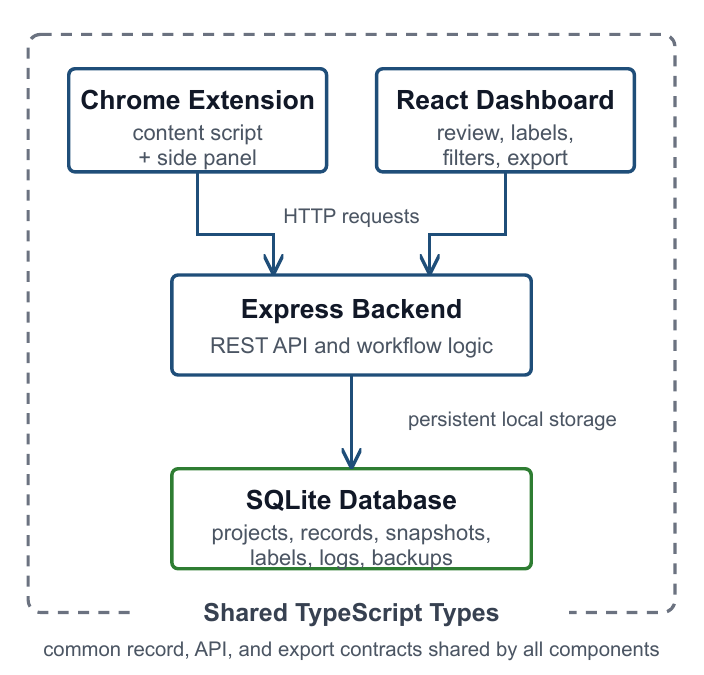}
  \caption{\toolname{} system architecture: a Chrome extension feeds
  browser-extracted evidence to an Express backend over a SQLite store, which
  serves a React dashboard for review, labeling, and export.}
  \label{fig:architecture}
\end{figure}

\toolname{} is implemented as a single TypeScript project with four parts. The
extension is a Manifest V3 Chrome extension with content-script extraction and a
side-panel UI\@. The backend is an Express application backed by SQLite through
\texttt{better-sqlite3}. The dashboard is a React application built with Vite. A
set of shared TypeScript types defines records, snapshots, comments, labels,
annotations, notes, reviews, exports, and backups. These types provide a common
contract for the extension, backend, and dashboard. Figure~\ref{fig:architecture}
illustrates the data flow: the extension extracts evidence from GitHub
pages and sends it to the Express backend, which persists records in SQLite and
serves the React dashboard used for review, labeling, and export.

The SQLite schema comprises roughly twenty tables covering projects, users,
records, snapshots, comments, research-label categories and labels, annotations,
notes, linked artifacts, fix evidence, record reviews, taxonomy and backfill
state, consensus labels, update-check runs and change logs, quality checks,
audit logs, and provenance. It enforces a project-scoped duplicate key over
repository owner, repository name, record type, and number.

\begin{table}[htbp]
\caption{\toolname{} implementation components.}
\label{tab:components}
\centering
\small
\begin{tabular}{>{\raggedright\arraybackslash}p{0.32\columnwidth}
                >{\raggedright\arraybackslash}p{0.58\columnwidth}}
\toprule
Component & Responsibility \\
\midrule
Chrome extension & Detects GitHub issue and pull-request pages, extracts
visible evidence, and drives side-panel collection. \\
Express backend & Serves the collection, review, refresh, export, backup, and
analysis APIs over local storage. \\
SQLite database & Stores projects, records, snapshots, comments, labels,
notes, reviews, audit logs, and update history. \\
React dashboard & Supports project management, filtering, labeling,
consensus review, quality checks, and export. \\
Shared types & Define the TypeScript contracts shared by the extension,
backend, dashboard, tests, and export schema. \\
\bottomrule
\end{tabular}
\end{table}

The storage layer emphasizes provenance and repeatability. Each collected record
can hold multiple snapshots, and a snapshot contains raw text, optional raw HTML,
normalized JSON, source method, capture time, and field provenance. The record
itself stores record-level fields: project, repository owner, repository name,
record type, number, title, state, collector, collection time, refresh time, and
update time. The project-scoped uniqueness constraint warns about duplicates
without preventing the same GitHub issue from being studied in different
projects.

All of these tables are included in JSON exports and project backups, so a
reviewer can inspect not only the final labels but also the evidence and
intermediate decisions that produced them. The fix-evidence fields in particular
are motivated by the long-running use of issue, change, and fix-link evidence in
software evolution studies \cite{borg2019szzunleashed}.

\section{Usage Demonstration}
\label{sec:demo}

This section demonstrates \toolname{} on a self-collected dataset rather than
reporting a controlled user study. We collected 20 issues from
\texttt{matplotlib/matplotlib} across two \toolname{} projects. The first,
\textit{Matplotlib Backend and Rendering Issue Evidence Study}, focuses on
backend behavior, rendering defects, documentation rendering problems, animation
behavior, maintenance discussions, and installation/security warnings. The
second, \textit{Matplotlib Regression and Compatibility Evidence Study}, focuses
on regression and compatibility evidence. For this project, we also injected
four synthetic multi-reviewer conflicts to exercise the consensus workflow. These
conflicts are not observations of real reviewer disagreement and serve only to
exercise how the dashboard surfaces conflicts. The demonstration does not
measure GitHub collaboration behavior broadly; instead, it exercises the kinds of
social and technical information that prior GitHub studies draw on
\cite{tsay2014influence,gousios2014exploratory}.

For each record, we verified that \toolname{} preserved the GitHub URL, title,
state, labels, issue body, captured comments, stored snapshot, screening
rationale, note, fix evidence, and research annotations. Across the two projects,
the validation notes report 20 records, 22 stored snapshots, 38 captured
comments, 20 research notes, 98 research annotations, 20 screening reviews, 20
fix-evidence entries, and the four injected unresolved consensus conflicts; the
artifact package includes the detailed record-by-record notes.

\begin{table}[htbp]
\caption{Validation dataset summary.}
\label{tab:validation}
\centering
\small
\begin{tabular}{>{\raggedright\arraybackslash}p{0.48\columnwidth}
                rrr}
\toprule
Measure & Backend & Regression & Total \\
\midrule
Records & 10 & 10 & 20 \\
Snapshots & 12 & 10 & 22 \\
Captured comments & 21 & 17 & 38 \\
Research notes & 10 & 10 & 20 \\
Research annotations & 60 & 38 & 98 \\
Screening reviews & 10 & 10 & 20 \\
Fix-evidence entries & 10 & 10 & 20 \\
Consensus conflicts & 0 & 4 & 4 \\
\bottomrule
\end{tabular}
\end{table}

The validation results in Table~\ref{tab:validation} indicate that \toolname{}
can preserve the main artifacts needed for a manually coded GitHub issue-labeling
study. The backend and rendering project exercises collection, screening,
labeling, and evidence capture. The regression and compatibility project shows
that the dashboard can surface disagreement cases, such as conflicting symptom
or component labels, without overwriting either reviewer's annotation.

The automated test suite comprises 37 passing tests covering extension URL
detection and extraction fixtures, attribute selection, database initialization,
backend collection and refresh APIs, duplicate handling, scoped exports, project
backup, consensus and disagreement workflows, audit logs, and dashboard
rendering. The suite includes scoped export checks for single-record and
filtered-record exports and backup round-trip checks confirming that records,
snapshots, comments, notes, annotations, and provenance are preserved.

\section{Tool Availability}

\toolname{} is open source. The source code is publicly hosted at
\url{https://github.com/t3-research/RepoTrace}. An archived, citable snapshot is
available at \url{https://doi.org/10.5281/zenodo.20954131}, and the demonstration
screencast is available at \url{https://youtu.be/ZEaeAzb2UkQ}. The artifact
includes the full source, setup and reproducibility instructions, the automated
test suite, seeded demo data, and validation notes.

The artifact runs locally and requires only Node.js, npm, SQLite, and a Chromium-based
browser such as Chrome. After \cmd{npm install}, a user can seed the
demonstration database (\cmd{npm run db:seed-demo}), run the automated test suite
(\cmd{npm test}), and build the dashboard and extension (\cmd{npm run build}).
The reproducibility instructions also cover API checks, project backup
import/export, update checks, and JSON export inspection, and explain how to
reproduce the validation dataset reported in
Section~\ref{sec:demo}.

\section{Conclusion}

\toolname{} demonstrates a browser-assisted workflow for constructing auditable
GitHub research datasets. By storing page evidence, source metadata, research
labels, notes, review decisions, refresh history, and exportable provenance in a
single local workspace, the tool makes the manual data-collection process easier
to inspect and reproduce. It supports the full local workflow---collection,
labeling, refresh, consensus review, backup, and JSON export---and the
20-record Matplotlib validation shows that it can preserve both the source
evidence and the research interpretation for later review. We plan to extend
\toolname{} with CSV and analysis-table export, tighter navigation from labels
to the exact page evidence, richer multi-reviewer workflows, and dataset
versioning. We will consider optional AI assistance only after establishing the
reliability of the existing non-AI evidence workflow.

\begin{acks}
This research was supported by use of the Nectar Research Cloud, a
collaborative Australian research platform supported by the NCRIS-funded
Australian Research Data Commons (ARDC). This research was supported by
the Summer and Winter Vacation Research Scholarship Program at Monash
University. Yongqiang Tian is the corresponding author.
\end{acks}

\bibliographystyle{ACM-Reference-Format}
\bibliography{references}

\end{document}